\documentstyle[epsfig]{aipproc}
\begin{document}
\title{Calculations of radio pulses from High Energy Showers}

\author{Jaime Alvarez-Mu\~niz}
\address{Bartol Research Institute, University of Delaware, Newark, DE 19711.\\
alvarez@bartol.udel.edu}

\author{Enrique Zas}
\address{Departamento de F\'\i sica de Part\'\i culas,\\
Universidad de Santiago de Compostela, E-15706 Santiago, Spain.\\
zas@fpaxp1.usc.es}
\maketitle




\begin{abstract}
In this article we review the progress made in understanding 
the main characteristics of coherent \v Cerenkov radiation induced 
by high energy showers in dense media. A specific code developed for 
this purpose is described because it took a significant part in this 
process. Subsequent approximations developed for the calculation of 
radio pulses from EeV showers are reviewed. Emphasis is given to the 
relation between the shower characteristics and different features of 
the corresponding radio emission. 

\end{abstract}

\section{INTRODUCTION}

It was about 40 years ago that Askar'yan proposed the detection of 
very high energy particles through the coherent radio emission from 
all the particles in a high energy shower \cite{askaryan}. 
Although radio emission from showers had been discussed before, 
Askar'yan noted that showers would develop an 
excess of electrons of order $10\%$ independently of shower energy and 
that such excess would lead to coherent radiation.   
Indeed if the emission is coherent the electric field scales as 
the excess charge and as the shower energy rises, the relative 
contribution of radio emission increases with respect to other 
wavelengths (say optical \v Cerenkov). Since the number of particles 
in a shower is proportional to shower energy, the power in 
the coherent signal scales with the square of the primary energy. This 
fact together with the technical simplicity of detectors makes the technique 
attractive for detecting very high energy showers. It is remarkable that many 
detection possibilities addressed in this meeting were already in 
Askar'yan original work. 

Clearly coherence demands that the wavelength of the radiation exceeds 
the size of the radiating region. As electromagnetic showers develop 
through successive pair production and bremsstrahlung interactions, the 
particle distributions of showers of a given energy are, to a very rough approximation, similar when described in radiation length units 
\footnote{Hadronic showers can be regarded as a jet of hadrons along  
shower axis which generates a chain of electromagnetic subshowers 
through neutral pion decays and have similar scales.}.
The radiation length of different materials in units of matter depth 
basically scales with the inverse of the atomic number, being 
$37$~g~cm$^{-2}$ for air and not too different for many other 
abundant materials that have been proposed for radio experiments such 
as water, salt and sand. Because of the medium density the 
scale of the showers can however be completely different. 
While air showers have sizes in the km range and coherence 
effects can be obtained at frequencies of order 10 MHz, for showers 
in ice the scale is a meter and coherence is kept to frequencies of 
order 10~GHz. 

Since the power associated to the \v Cerenkov emission increases with frequency 
Askar'yan suggested detecting showers in dense medium, as those that 
could be induced by deeply penetrating particles underground or by 
cosmic rays on the Moon surface. 
Attempts were made to detect radio pulses in coincidence with other 
air shower detectors \cite{weekes} and, although the results were positive, 
it became clear that there were many difficulties. 
Calculations revealed that several additional mechanisms competed with 
excess charge in the generation of radio pulses, including dipole radiation 
and transverse currents because of the magnetic field of the Earth, 
and transition radiation when the shower reached the ground \cite{kahn}. 
The interpretation of radio signals was obscured by these and other 
difficulties and the technique was almost completely abandoned but for 
a few isolated efforts that have been reviewed in this conference. 

In the mid 1980's the efforts to build high energy neutrino telescopes 
renewed the activity in the radio technique \cite{markov,ralston} 
as a possible alternative to the projects to detect the \v Cerenkov light 
from muons under water or ice \cite{halzen}. 
As a result early in the 1990's a Monte Carlo program was developed for the 
calculation of distant radio signals from TeV electromagnetic showers in 
ice \cite{zhs91,zhs92}. These led to several efforts to assess the 
possibility  
to detect neutrino events with arrays of antennas in Antarctica 
\cite{provorov,frichter,jelley}. Later on a series of approximations were made 
to extend the results of these calculations to both electromagnetic and 
hadronic showers of energies in the EeV and ZeV 
ranges \cite{alz97,alz98,alz99,alz00}. 
In this article we will concentrate on the efforts made towards 
understanding \v Cerenkov radio emission in dense media from the 
excess charge that develops in high energy showers. 
The recent experimental confirmation of this effect in a 
SLAC experiment \cite{gorham}, in agreement with expectations, leaves now 
little doubt about the enormous potential of this technique. 
We believe that many developments will follow this conference and it 
is our hope that these calculational efforts will help in the task. 

\section{ZHS: A shower generator for radio emission.}

The calculation of radio pulses from showers is a complex problem in 
which the emission from all particles in a high energy shower have to 
be carefully added in a coherent fashion, taking into account the 
velocity and direction, position and time of each of the particles 
involved. For a medium such as ice, having a refraction index $n=1.78$ 
for radio waves, electrons of energy exceeding 107~keV emit 
\v Cerenkov radiation. The simulation 
of the shower cascade from this perspective was tackled by 
Zas, Halzen and Stanev in \cite{zhs92} with a result that has become 
known as the {\sl ZHS} Monte Carlo. Since the largest 
contribution to the excess charge in a shower is from electrons, 
only electromagnetic showers were considered. This program has been 
a reference point for much progress made during the 1990's in 
radio detection and deserves a brief description. Efforts have 
been made by other groups to test the code against other shower 
codes \cite{besson}. 

The longitudinal development of an electromagnetic shower mostly 
follows from successive pair production and bremsstrahlung interactions 
with the electric field of atomic nuclei that are the heart of the 
showering process. 
For both these processes the effective interaction distance increases 
as the energy of the interacting photon or electron rises and atomic 
screening of the atoms becomes progressively more important. 
In this respect the Landau-Pomeranchuck-Migdal (LPM) effect can be 
regarded as the higher energy extreme of this situation in which the 
atomic potential of the individual atoms has to be modified because 
of collective effects of several nuclei. 
The lateral structure of the shower is mostly controlled by elastic 
scattering. Because of the long range of the Coulomb interaction it 
turns out that most angular deviations arise  because of multiple 
elastic scattering which is treated in this Monte Carlo keeping the 
first two terms in Moli\`ere's expansion \cite{moliere}. 

Besides these interactions that are common to all simulation programs, 
the ZHS code has special features for radio calculations. 
It is a three 
dimensional code that takes timing of the particles in great detail. 
Time delays are measured with respect to a plane perpendicular to shower 
axis that starts in phase with the primary particle and moves parallel 
to the axis at speed $c$. 
Most of such delays are due to angular deviations, to 
multiple elastic scattering and to the subluminal velocities 
of the shower particles all of which are taken into account. 
The program 
includes electron positron annihilation, M\"oller, Bhabha and Compton scattering, which only become relevant at low energies but are 
responsible for the excess charge in the shower. These routines were 
developed following the EGS4 code \cite{egs4}. 
In the first version \cite{zhs91} it did not include the LPM effect but 
this was soon remedied \cite{zhs92}. 
It can be run with threshold energies as low as 100~keV, but 
its accuracy is expected to decrease for 
thresholds below this value because it does not include 
the photoelectric effect. 
Finally it is designed to be fast using, when possible, numerical parameterizations instead of actual functions to minimize calculating time. 

\begin{figure}[hb] 
\center{
\centerline{\epsfig{file=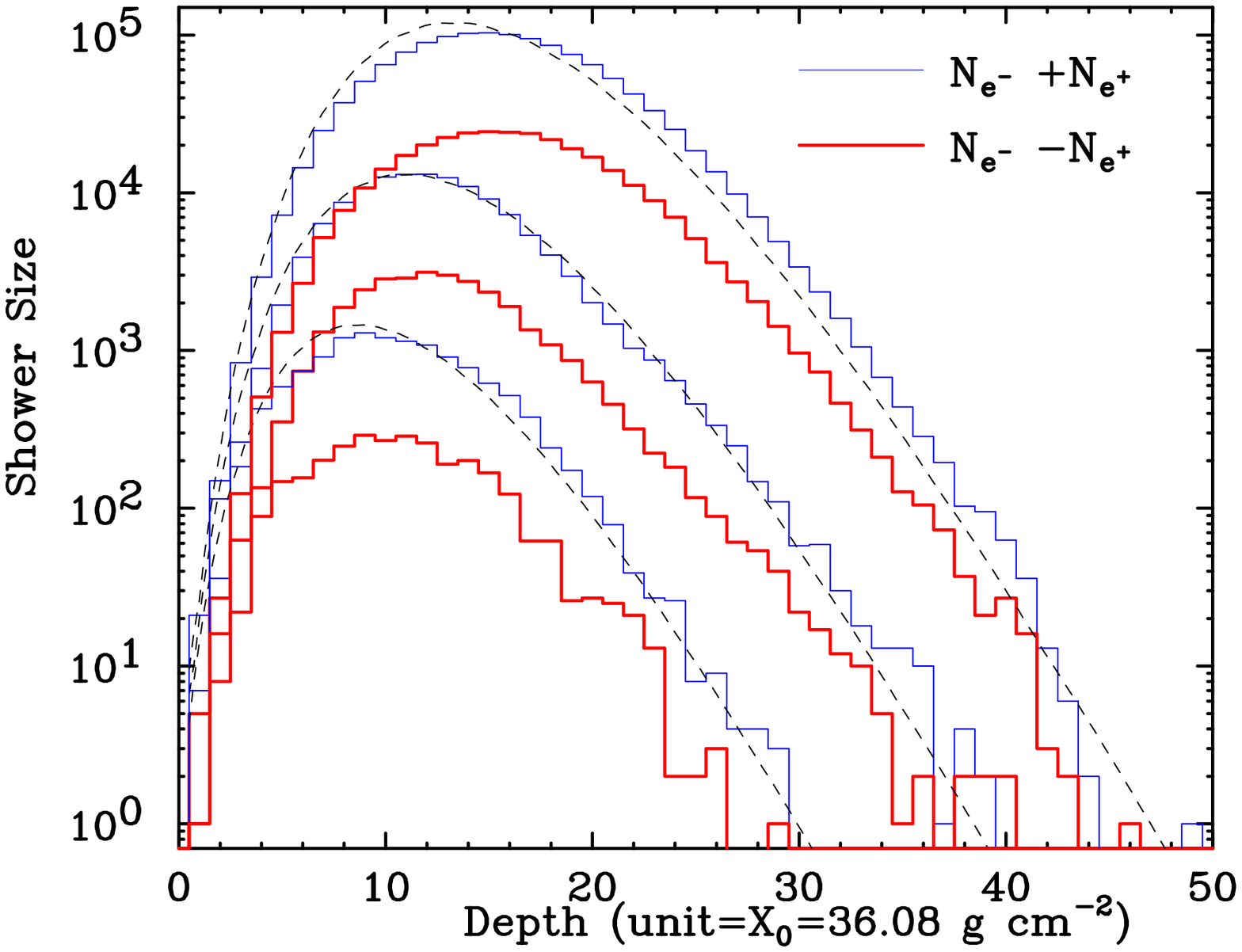,height=2.in,width=2.9in} 
\epsfig{file=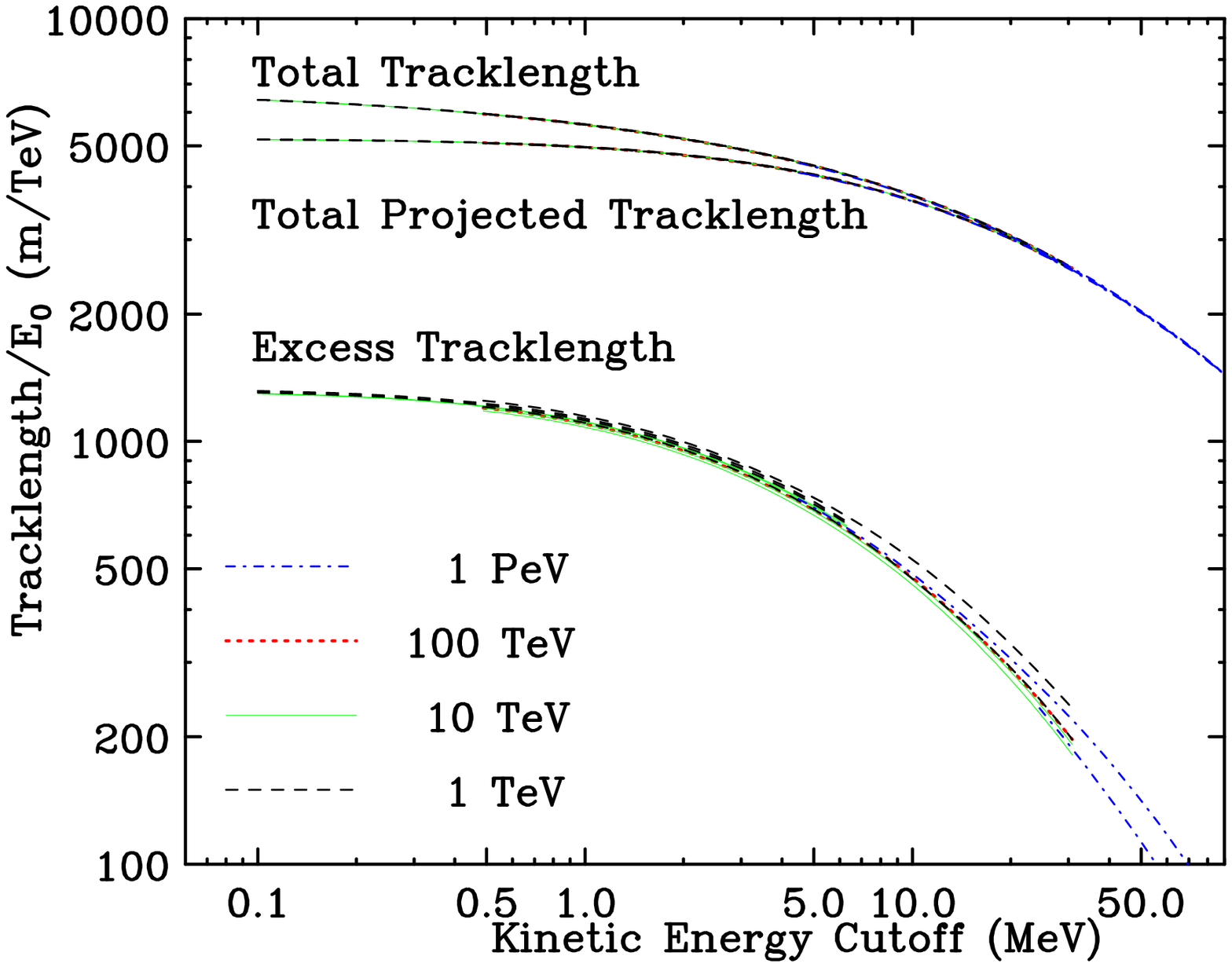,height=2.in,width=2.9in}}  }
\vspace{10pt}
\caption{{\bf a)} Left: 
Depth development of the total number of electrons and positrons and the
excess number of electrons for single 1, 10 and 100 TeV electron 
initiated showers.
The Monte Carlo threshold energy is 1 MeV. Depth is measured in radiation 
lengths of ice. The dashed lines are the results in approximation-B of shower
theory in the limit of $E_{th} \rightarrow m_e$.
{\bf b)} Right: 
Shower track lengths for all charged particles in 1, 10, 100~TeV and 1~PeV 
electron showers as a function of the calculational cutoff $(E_{th})$. The
three sets of curves correspond to i) the sum of all track lengths (Total 
Tracklength), ii) the sum of all track lengths projected onto the shower axis 
(Total Projected Tracklength) and iii) the difference of electron and
positron track lengths projected onto the shower axis (Excess Tracklength). 
Several showers are plotted for each energy to illustrate the small  
effect of fluctuations. The tracklengths have been normalized to the 
shower energy in TeV.}
\label{excess}
\end{figure}
The results obtained for the longitudinal development of the showers 
are consistent with the Greisen parameterization. The lateral distributions 
are consistent with three dimensional theoretical shower results provided 
the Moli\`ere radius is divided by a factor of about 2, in agreement with 
other simulations \cite{hillas}. The excess 
charge obtained is of order $20 \%$ depending on particle threshold, 
it is only slightly dependent on shower depth, see Fig.~\ref{excess}a.  
The extra track due to electrons is $21\%$ of the total 
(the \v Cerenkov emission is proportional to the tracklengths of the 
charged particles) as shown in Fig.~\ref{excess}b. 
Compton scattering of atomic electrons that 
{\sl get accelerated} into the shower account for most of the effect 
($12 \%$). It is remarkable that close to half ($14 \%$) of the excess 
tracklength is due to particles below 5~MeV (1~MeV). The calculation of 
coherent signals requires a 
lower threshold than implied by these numbers because the lower energy 
particles are expected to interfere more destructively as they 
get longer delays and larger angular deviations. 
This is already important because it sets an  
upper bound to the energy threshold necessary for an accurate calculation.

\section{\v Cerenkov radiation} 

The detection of radio pulses from air showers is clearly a very 
different problem from detecting them in a dense medium such as ice.  
Firstly in ice the {\sl coherence band} extends to the GHz regime.
Secondly most shower particles travel very small paths of tens of 
centimeters, and they are hardly deviated because of the Earth's 
magnetic field. As a result the excess charge mechanism is expected 
to dominate the radiation and the problem becomes simpler from the 
calculation point of view. 
Moreover there is a substantial difference in the relation between 
the typical observation distances and the shower dimensions. 
While for air showers the observation distance is of the some order or 
even less than the shower dimensions, prospects of detecting radio pulses 
under the Earth surface or from the Moon are attractive precisely because 
they can be detected from very large distances compared with the shower 
dimensions which are of order one meter. For most interesting cases the 
Fraunhofer approximation turns out to be accurate what is 
useful for calculational purposes and simplifies both the interpretation 
and the presentation of results. 

When a particle of charge $z$ moves through a medium of refractive index $n$ 
with velocity $\vert \vec {v} \vert=\beta c > c/n$ \v Cerenkov light is 
emitted at the \v Cerenkov angle $\theta_C$, verifying 
$\cos \theta_C=(\beta n)^{-1}$, with a power 
spectrum given by the well known Frank-Tamm result \cite{fran} : 
\begin{equation}
{d^2 W \over  d\nu d l} = 
\left[{4\pi^2\hbar\over c}\,\alpha\right]z^2\nu
\left[1-{1\over\beta^2 n^2}\right]\;, 
\label{tamm}
\end{equation}
with $\nu$ the frequency, $c$ the speed of light, $d l= c \beta d t$ a 
small element of particle track length, and $\alpha$ the fine structure 
constant. This is the standard approximation used for most 
\v Cerenkov applications for wavelengths orders of magnitude
smaller than the tracks. 
This expression corresponds to an infinite track and  
in the case of a finite track there are diffraction effects. 

Working with the time Fourier transform of the electric field, 
$\vec{E}(\omega)$, Maxwell's equations can be solved in the 
transverse gauge for an arbitrary current \cite{zhs92}:
\begin{equation}
{\vec E}(\omega, {\vec {\rm x}})= 
{{\rm e} \mu_{\rm r} \over 2 \pi \epsilon_0 {\rm c}^2}
i \omega \int \int \int \int {\rm dt'~d}^3 {\vec {\rm x'}} ~
{\rm e}^{i \omega {\rm t'}+ i \vert {\vec k} \vert 
          \vert {\vec {\rm x}} - {\vec {\rm x'}} \vert} 
~{\rm { {\vec J}_{\perp}(t',{\vec {x'}}) \over 
                \vert {\vec {\rm x}} - {\vec {\rm x'}} \vert } } 
\label{generalsol}
\end{equation}
where $\vert \vec k \vert=k=n \omega / \rm c$ and ${\vec {\rm J}}_{\perp}$ 
is a divergenceless component of the current transverse to the observation 
direction, $\mu_r$ is the relative permeability of the medium and 
$\epsilon_0$ the permittivity of the vacuum. Eq.~\ref{tamm} can be 
obtained from this expression. For a
single particle moving uniformly with velocity $\vec v$ between 
times $t_1$ and $t_1+\delta t$ in the Fraunhofer limit it simplifies to: 
\begin{equation}
{\vec E}(\omega,{\vec {\rm x}})=
{e \mu_{\rm r}~i \omega \over 2 \pi \epsilon_0 {\rm c}^2}~
{{\rm e}^{i k R } \over R} ~ 
{\rm e}^{i(\omega - \vec k \cdot \vec v) {\rm t}_1}~ 
{\vec v}_{\perp }~ 
\left[{{\rm e}^{i(\omega - \vec k \cdot \vec v) \delta {\rm t}} -1 
                        \over i (\omega - \vec k \cdot \vec v)} \right]
\label{algorithm}
\end{equation}

The Fraunhofer limit corresponds to an observation distance 
$R >> [v \delta t sin \theta]^2/ \lambda$. For a characteristic track 
of one radiation length ($\sim 40$~cm) and $\lambda$ of 30~cm corresponding 
to 1~GHz frequencies, the condition implies $R>> 50$~cm which is well 
satisfied for detection distances of order one km. 
It should be remarked that the inequality can be always satisfied 
provided a sufficiently small time interval is chosen. 
The ZHS program uses this expression. For an infinitesimal track 
$\delta l$ it becomes: 
\begin{equation}
R \vec E(\omega,{\vec {\rm x}})=
{e \mu_{\rm r}~i \omega \over 2 \pi \epsilon_0 {\rm c}^2}~
{\vec  \delta l}_{\perp} ~ 
{\rm e}^{i(\omega -{\vec k} {\vec v}_1){\rm t_1} } 
~{\rm e}^{ikR},
\label{field} 
\end{equation}
This expression shows that the electric field amplitude is 
proportional to the tracklength and that radiation is polarized 
in the direction of $\vec l_{\perp}$, the apparent direction 
of the track as seen by the observer.  

\begin{figure}[hb] 
\center{
\centerline{\epsfig{file=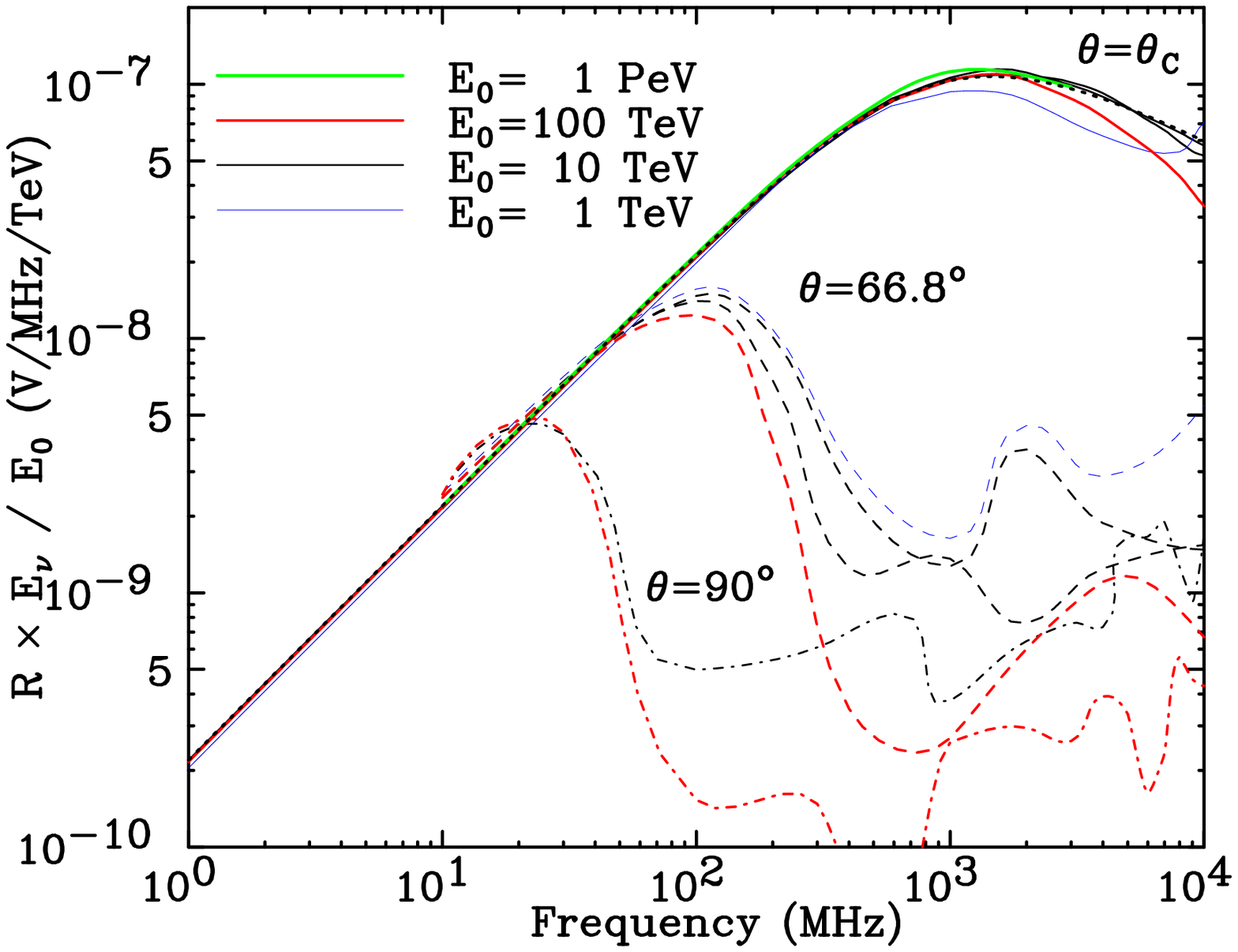,height=2.in,width=2.9in} 
\epsfig{file=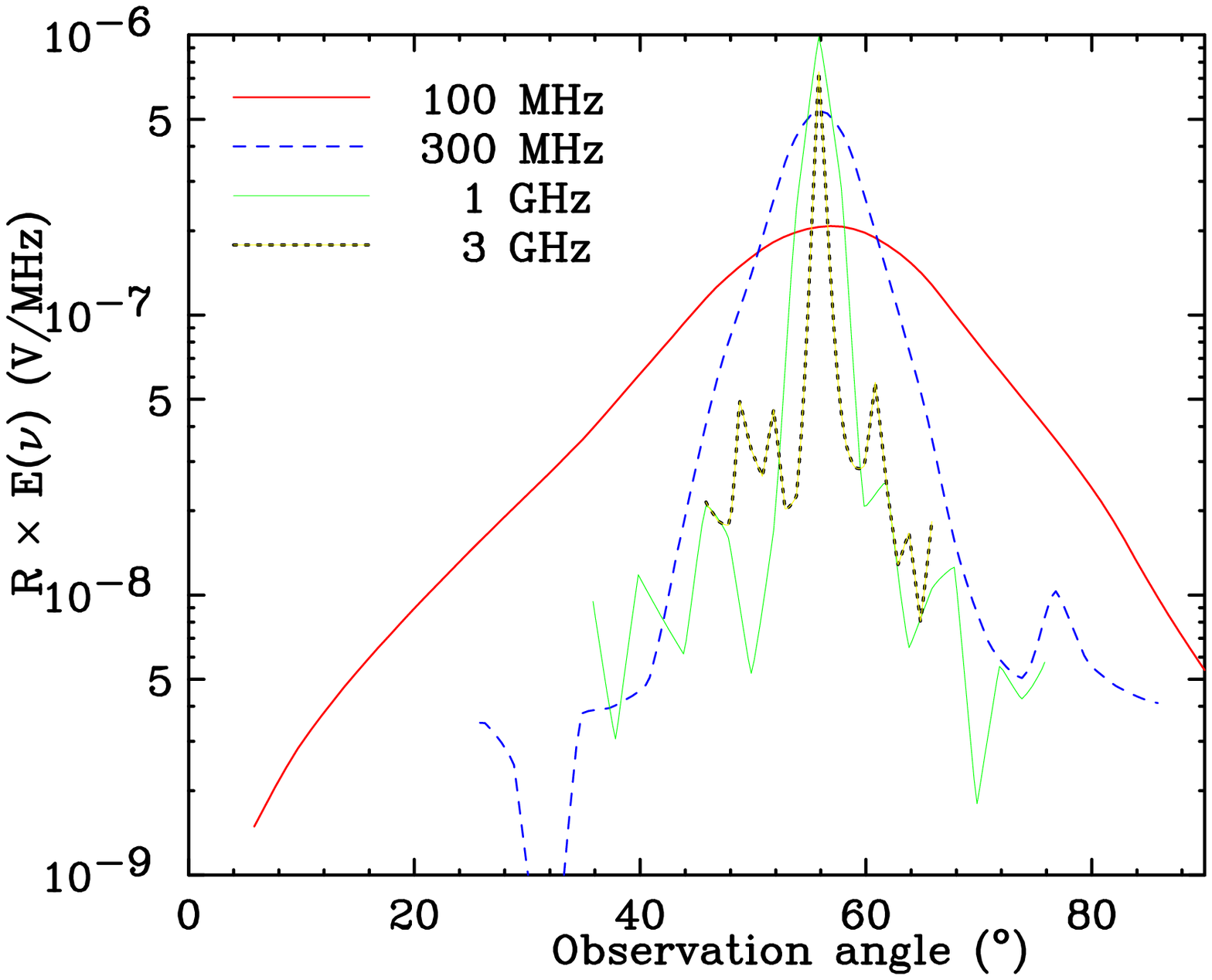,height=2.in,width=2.9in}}  }
\vspace{10pt}
\caption{{\bf a}) Left: 
Frequency spectrum of the electromagnetic pulse. The figure displays 1, 
10 and 100~TeV showers for different thresholds. The amplitudes have been
divided by the energy of the primary in TeV and renormalized depending on the
threshold. The full curves correspond to observation at the \v Cerenkov angle 
$(\theta_{C})$ to the shower axis. The dashed (dot-dashed) curve corresponds
to observation at $66.8^{\circ}$ (at $90^{\circ}$) to the same axis.
{\bf b}) Right: 
Angular distribution of the electric field generated by a 10~TeV shower 
($E_{th}=611$~keV). The observation angle is the polar angle of the 
radiation with respect to the shower axis.}
\label{amplitude}
\end{figure}

For a high energy shower the calculation of the radio pulse is now a 
matter of book-keeping. 
Results of the electric field amplitude in the \v Cerenkov direction 
display a characteristic frequency spectrum: a linear 
rise with frequency until the GHz region where destructive interference 
sets in and the spectrum gradually levels off as shown in 
Fig.~\ref{amplitude}a. Although 
in the \v Cerenkov direction all points of the trajectory of a particle 
moving at the speed of light along the shower axis are emitting in phase, 
lateral deviations of the particle trajectories and time delays 
destroy the coherent interference at sufficiently high frequencies. 
The effect of the lateral distribution is dominant.  
The amplitude at a given frequency is proportional to the excess 
tracklength which in turn accurately scales with shower energy as 
shown in Fig.~\ref{excess}b. 

The polar diagram of the emitted radiation displays a characteristic 
peak in the \v Cerenkov direction of a width which is inversely 
proportional to the frequency. This is plotted in Fig.~\ref{amplitude}b 
and corresponds to a diffraction-like pattern. The width of the 
diffraction peak about the \v Cerenkov angle depends on the 
longitudinal distribution of the charge in the 
shower. As the observation angle gets away from the 
\v Cerenkov direction different longitudinal sections of the shower 
development start to interfere destructively and the amplitude of the 
electric field drops. 

\begin{figure}[hbt]
\centering
\centerline{\epsfig{file=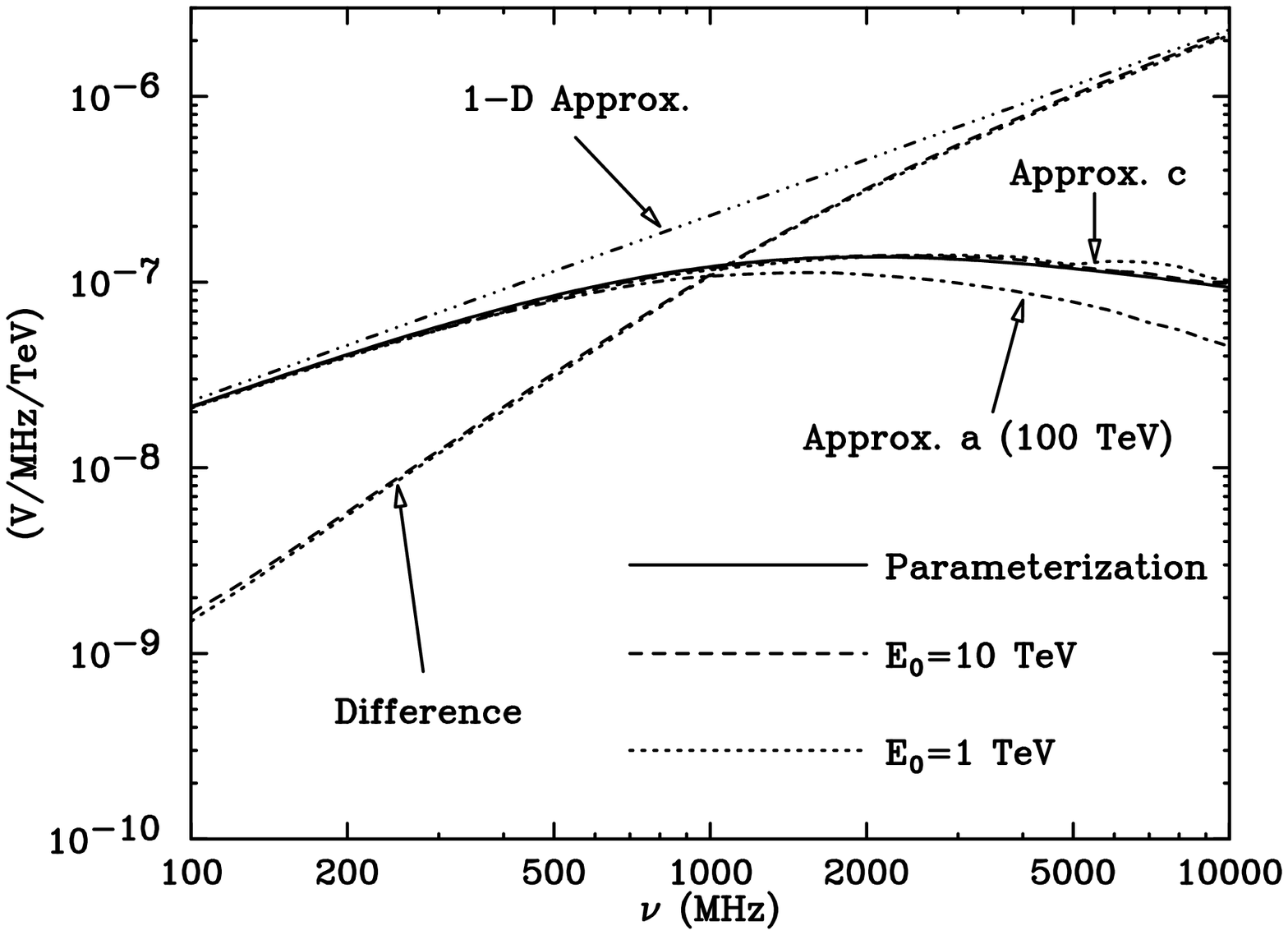,height=3.3in,width=4.5in} }
\caption{Comparison of 
complete simulation results for the frequency spectrum in the 
\v Cerenkov direction for 1 and 10~TeV electromagnetic showers in 
approximation $c$, for a 100~TeV in the standard approximation 
($a$) and with the 1D-approximation (top curve). 
The lower curves represent the difference between the 1D approximation 
and the full simulation results using in the $c$ approximation. 
Note that both the spectrum and the difference have the same behavior 
for different shower energies. 
All radio pulses scale with shower energy and are normalized to 1~TeV. 
}
\label{diff-freq}
\end{figure}

Long tracks in a shower can be subdivided in 
arbitrary subintervals. Provided the Fraunhofer approximation is valid 
for the shower as a whole, the algorithm of Eq.~\ref{algorithm} 
for a particle that moves at constant speed displays cancellations 
between all the end points of the intermediate subintervals of the 
track. Only the end points of the total track need to be considered 
in this case what simplifies the computational task enormously. 
In approximation ``a" (default) each electron gives rise to a single 
track to which Eq.~\ref{algorithm} is applied. 
As particles do suffer some deceleration along the track the average 
velocity is used. Different criteria for this subdivision have given 
rise to approximations ``b" and ``c" \cite{rome}. 
The approximation ``a" has been shown to provide 
sufficiently good results when compared to longer calculations 
in which each electron track is subdivided into subtracks each of which contributing with Eq.~\ref{algorithm} as shown in Fig.~\ref{diff-freq}. 
Note that approximation ``a" 
is conservative in the sense that it underpredicts the amplitude of 
the pulse at very high frequencies. 

\section{1-D Approximation}

With the help of scaling it is possible to raise the shower simulation 
threshold and to generate showers of about 1~PeV but at higher energies 
the calculations become too lengthy to be handled. The calculation of 
radioemission from EeV showers is however of great interest because 
the technique is expected to be most competitive for these energies. 
Because of the LPM suppression of bremstrahlung and pair production, 
these showers have very different elongated depth distributions. 
As a result the angular width of the \v Cerenkov peaks is expected to 
be reduced with respect to showers of energy below 1~PeV. 
As the lateral distributions of these showers are not expected to be 
very different from the lower energy showers, not many effects can be 
anticipated in the frequency spectrum of the signal in the 
\v Cerenkov direction. 

The one dimensional approximation has been developed to study the 
effects of the longitudinal elongation \cite{alz97,alz98}. 
In this approximation the particles in the shower are all assumed 
to be moving at the speed of light and the lateral distribution is 
neglected. 
The current density simply becomes \cite{alz00}: 
\begin{equation}
{\vec J}_\perp (\vec {\rm x'},{\rm t'})=Q(z')~
{\vec c}_\perp ~\delta^3 (\vec {\rm x'}-{\hat n_{z}}ct')
\label{current}
\end{equation}
where the shower develops along the $z$ axis, ${\hat n_{z}}$ is a unit 
vector in this direction and $Q(z')$ is the 
excess charge in the shower. The substitution of this current into 
Eq.~\ref{generalsol} leads to:
\begin{equation}
{\vec E}(\omega,{\vec {\rm x}})=
{e \mu_{\rm r}~\over 2 \pi \epsilon_0 {\rm c}^2}~i {\omega\over c}
~\sin\theta~{\hat n_\perp}~\int dz'~ Q(z')~{e^{i{\omega\over c} z'+
ik\vert {\vec {\rm x}}-
z'{\hat n_{z}}\vert}\over \vert {\vec {\rm x}}-z'{\hat n_{z}} \vert}
\label{fresnel}
\end{equation}
where $\theta$ is the angle between the shower axis and the direction 
of observation ${\vec {\rm x}}$ and ${\hat n_\perp}$ is a unitary vector
perpendicular to ${\vec {\rm x}}$. It should be stressed that 
no further approximations are made at this stage.

The phase factor in Eq.~\ref{fresnel}
can be approximated by $ik\vert {\vec x}-{\vec z}~'\vert \simeq 
ik R - i {\vec k}\dot {\vec z}~'$ in the Fraunhofer limit. Here 
$R=\vert {\vec {\rm x}} \vert$ is 
the distance from the center of the shower to the observation point. 
The electric field amplitude simply becomes the Fourier transform of 
the longitudinal charge distribution:
\begin{equation}
\vec E(\omega,{\vec {\rm x}}) =
{e \mu_{\rm r}~\over 2 \pi \epsilon_0 {\rm c}^2}~i \omega
~\sin \theta~{{\rm e}^{ikR}\over R}~{\hat n_\perp}
\int dz'~Q(z')~{\rm e}^{i p z'}
\label{fraunhofer}
\end{equation}
where we have introduced for convenience the parameter
$p(\theta,\omega)= (1-n \cos \theta)~\omega /c$ in Eq.~\ref{fraunhofer}. 
The angular pattern around the \v Cerenkov direction is precisely the 
analog of the classical diffraction pattern of an aperture function. 

The accuracy of the approximation has been studied comparing it to 
complete simulations. 
Excellent agreement is obtained for frequencies below 100~MHz. For  
frequencies progressively increasing above this value, the electric 
field becomes overestimated in the \v Cerenkov direction, but the 
angular distribution of the pulse is otherwise preserved. 
This is not surprising since the lateral distribution is ignored 
and no levelling off in the frequency spectrum can be expected. 
Moreover the difference between the complete simulation and the 
approximation scales with shower energy as shown in 
Fig.~\ref{fraunhofer}.  
Ad hoc corrections can be implemented to make quantitative predictions 
in the GHz regime accounting in an effective way for the ignored effects 
\cite{alz98,alz00}. 
This is analogous to the form factor described in \cite{buniy}. 
The approximation gives some insight into the complexity of the 
spectrum and angular distribution of the pulses relating them 
to features in shower development. 

\begin{figure}[hbt]
\centering
\mbox{\epsfig{figure=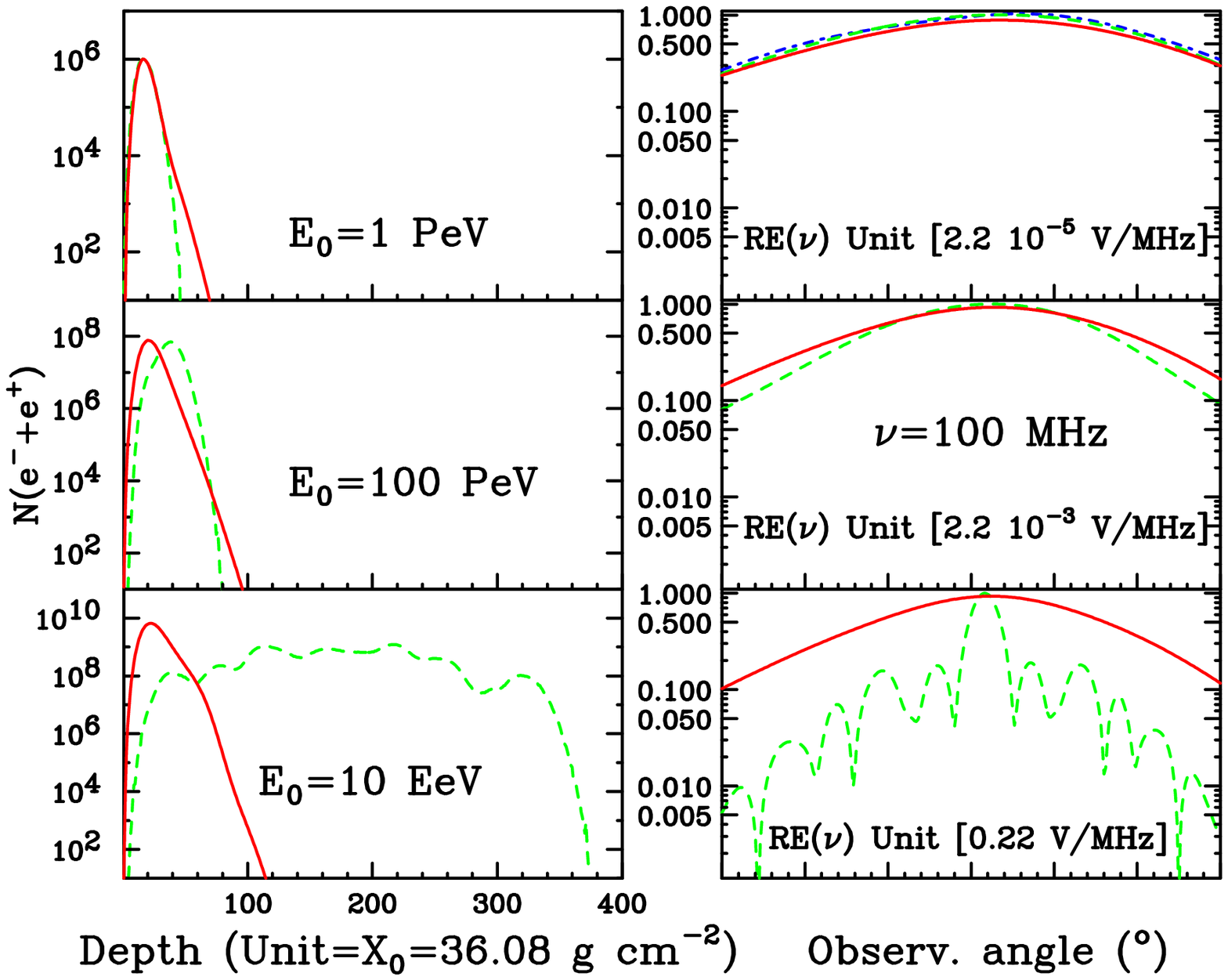,width=5.8in}}
\caption{Left: Longitudinal development of electromagnetic 
(solid red lines) and hadronic (dashed lines) showers in ice  
for different energies. Right: Angular distribution of the electric 
field amplitude times the distance emitted by the showers shown on 
the left. Shown is the value $| E(\nu) \, R |$ where $R$ is the 
distance to the shower, normalized to its maximum at the Cherenkov 
angle ($\theta_C= 56^\circ$). The units, that is the 
precise values at maximum are marked in the figure.
For the 1 PeV case the electric field amplitude obtained in  
a complete simulation (dot-dashed line) is also shown to 
compare with the 1D approximation. }
\label{fig:prd1emhad}
\end{figure}

The development of EeV showers of electromagnetic \cite{alz97} and 
hadronic \cite{alz98} character has been studied using hybrid techniques 
that combine simulation and parameterizations of low energy showers. 
The LPM suppression leads to a very large elongation of electromagnetic 
showers which results in a reduction of the angular width of the 
diffracted pulses. The elongation has been calculated to scale as 
$E_{sh}^{1/3}$ for shower energies satisfying $E_{sh} > 20~$PeV \cite{alz97}. 
For hadronic showers the elongation does not affect shower development 
significantly until energies in the EeV range. This is because high 
energy hadronic interactions have large multiplicities and the particles 
that emerge carry small fractions of the primary energy. Moreover the 
neutral pions at very high energies interact before decaying into photons, 
and the transfer of energy to electromagnetic particles is postponed until 
the pions have energies in the few 10's of PeV. The emerging photons 
induce ordinary showers. Photons or electrons from very short lived 
particles are the main source of the occasional shower elongations observed 
\cite{alz98}.  
This is illustrated in Fig.~\ref{fig:prd1emhad}.

Eq.~\ref{fresnel} can be also used for calculating the radio 
emission in an approximate fashion to establish Fresnel interference 
effects. Since this approximation neglects the lateral distribution 
the results are only expected to be valid for showers 
which are very elongated because of the LPM effect. For these 
showers the Fresnel interference effect becomes more important than 
the lateral distributions. 
As the observation distance decreases the angular distribution of the 
pulses becomes wider and the main peak drops. 
The results show that at the Fresnel distance 
defined as $R_F=\pi n ~\nu ~ (L_s \sin\theta/2)^2 /c$ 
($L_s$ is the shower length) deviations of the electric field in the 
\v Cerenkov direction from the Fraunhofer approximation are below $10 \%$. 

For neutrino induced showers the two types of showers become relevant. 
Neutrinos always induce a hadronic character shower 
through the nucleon fragments. In the case of charged current electron 
neutrino interactions, the emerging shower has mixed hadronic and 
electromagnetic character and the relative weights of each component 
depend on $y$, the fraction of energy transferred to the nuclear fragments. 
For very high energy neutrinos this leads to an interesting combination 
of two showers of different lengths because of the relative importance 
of the LPM effect for both types of showers. The resulting angular 
distributions of the radio pulse displays a complex pattern 
of the two corresponding angular widths. It has been shown that it is in 
principle possible to use the angular information of the radio pulse to 
extract the corresponding value of $y$ for the interaction \cite{alz99}. 

\section{Summary}

We have reviewed the development in calculations of the full diffraction 
patterns and frequency spectra of hadronic and electromagnetic showers 
developing in ice. The complex emerging patterns are well 
understood in terms of the shower properties. 
We have stressed the dependence on the angular 
width of the \v Cerenkov pulse on length and its implication for 
electromagnetic showers with strong LPM elongations. 
 
\section*{Acknowledgements}

We thank D.~Saltzberg for helpful comments after carefully reading the manuscript and the organizers P.~Gorham and D.~Saltzberg for providing 
this unique oportunity to give a good impulse to this field. 
This work was supported in part by the European 
Science Foundation (Neutrino Astrophysics Network N. 86), by the CICYT (AEN99-0589-C02-02) and by Xunta de Galicia (PGIDT00PXI20615PR).
The research activities of J.A-M at Bartol Research Institute
are supported by the NASA grant NAG5--7009.


\begin{thebibliography}{9}
%
%
\bibitem{askaryan} G.A. Askar'yan, Zh. Eksp. Teor. Fiz {\bf 41}, 616 (1961)
[Soviet Physics JETP {\bf 14} 441, (1962)]; {\bf 48}, 988 (1965) [{\bf 21},
658 (1965)].
%
\bibitem{weekes} T. Weekes, in these proceedings and references therein.   
%
\bibitem{kahn} F.D.~Kahn and I.~Lerche, {\sl Proc. of the Roy. Soc. A} 
{\bf 289}, 206 (1966).   
%
\bibitem{markov} M.A. Markov, I.M. Zheleznykh,
Nucl. Instr. and Methods Phys. Res. {\bf A248}, 242 (1986).
%
\bibitem{ralston} Ralston,~J.P. and McKay,~D.M.,  {\sl Proc.\ Astrophysics in
Antarctica Conference}, ed.\ Mullan,~D.J.,  Pomerantz,~M.A. and
Stanev,~T., (American Institute of Physics, New York, 1989) Vol.~198, p.~241%
\bibitem{mannheim} K. Mannheim, Astropart. Phys. {\bf 3}, 295 (1995).   
%
\bibitem{halzen} F.~Halzen, in these proceedings.
%
\bibitem{zhs91} F.~Halzen, E.~Zas, and T.~Stanev, {\sl Phys.\ Lett.} 
{\bf B257} 432 (1991).
%
\bibitem{zhs92} E.~Zas, F.~Halzen, and T.~Stanev, 
{\sl Phys. \ Rev.} {\bf D45}, 362 (1992).
%
\bibitem{provorov} A.L. Provorov, I.M. Zheleznykh, Astroparticle Physics 
{\bf 4}, 55 (1995).
%
\bibitem{frichter} G.M. Frichter, J.P. Ralston, D.W. Mc Kay,
{\sl Phys. \ Rev.} {\bf D 53}, 1684 (1996).
%
\bibitem{jelley} J.V.~Jelley, {\sl Astropart. Phys.} {\bf 5}, 255 (1996).
%
\bibitem{alz97} J.~Alvarez-Mu\~niz and E. Zas, {\sl Phys.\ Lett.} {\bf B411}, 218 (1997).
%
\bibitem{alz98} J.~Alvarez-Mu\~niz and E. Zas, {\sl Phys.\ Lett.} {\bf B434}, 396 (1998).
%
\bibitem{alz99} J.~Alvarez-Mu\~niz, R.A.~V\'azquez and E.~Zas, 
{\sl Phys. \ Rev.}  {\bf D 61}, 023001 (2000).
%
\bibitem{alz00} J.~Alvarez-Mu\~niz, R.A.~V\'azquez, E.~Zas, 
{\sl Phys. \ Rev.}  {\bf D 61}, 023001 (2000).
%
\bibitem{gorham} D.~Saltzberg in these proceedings, P.~Gorham, {\sl et. al},  
e-print archive: hep-ex/0011001.
%
\bibitem{besson} D.~Besson {\sl et al.}, in these proceedings.
%
\bibitem{moliere} Moli\`ere,~G., {\sl Z.\ Naturforsch} 
{\bf 3a} 78 (1948).
%
\bibitem{egs4} Nelson,~W.R., Hirayama,~H. and Rogers,~D.W.O., 
{\sl The EGS4 Code System},  SLAC Rep.~SLAC-265, UC-32, 
Stanford Linear Accelerator Center (1985).
%
\bibitem{hillas} M.~Hillas, private communication. 
%
\bibitem{fran} Frank,~I. and Tamm,~I., {\sl Dokl.\ Akad.\ Nauk\ SSSR} {\bf 14}
 109 (1937).
%
\bibitem{rome} J.~Alvarez-Mu\~niz, G.~Parente, and E.~Zas, {\sl Proc.\ XXIV
Int. Cosmic Ray Conf.}, Rome (1995), vol 1, p. 1023.
%
\bibitem{buniy} R. Buniy, J.P. Ralston, astro-ph/0003408. 
%
\end{thebibliography}
\end{document}